\documentclass{kluwer}
\usepackage{graphicx}

\begin{document}
\begin{article}
\begin{opening}
\title{DETERMINATION OF ELECTRON FLUX SPECTRA IN A SOLAR FLARE WITH AN
AUGMENTED REGULARIZATION METHOD: APPLICATION TO RHESSI DATA}


\author{Eduard P. \surname{Kontar}\thanks{Off-print requests: {\tt eduard@astro.gla.ac.uk}}}
\institute{Department of Physics \& Astronomy, University of Glasgow, G12 8QQ, UK}
\author{A. Gordon \surname{Emslie}}
\institute{Department of Physics, The University of Alabama in Huntsville, Huntsville, AL 35899,
USA}
\author{Michele \surname{Piana}}
\institute{Dipartimento di Matematica, Universit\`a di Genova, via
Dodecaneso 35, I-16146 Genova, Italy}
\author{Anna Maria \surname{Massone}}
\institute{INFM, UdR di Genova, via Dodecaneso 33, I-16146 Genova,
Italy}
\author{John C. \surname{Brown}}
\institute{Department of Physics \& Astronomy, University of Glasgow, G12 8QQ, UK}




\runningtitle{Determination of Electron Spectra in Solar Flares}
\runningauthor{Kontar et al.}

\begin{abstract}
Kontar et al. (2004) have shown how to recover mean source electron spectra ${\overline F}(E)$ in
solar flares through a physical constraint regularization analysis of the bremsstrahlung photon
spectra $I(\epsilon)$ that they produce. They emphasize the use of non-square inversion
techniques, and preconditioning combined with physical properties of the spectra to achieve the
most meaningful solution to the problem. Higher-order regularization techniques may be used to
generate ${\overline F}(E)$ forms with certain desirable properties (e.g., higher order
derivatives). They further note that such analyses may be used to infer properties of the electron
energy spectra  at energies well above the maximum photon energy observed. In this paper we apply
these techniques to data from a solar flare observed by {\it RHESSI} on 26 February, 2002. Results
using different orders of regularization are presented and compared for various time intervals.
Clear evidence is presented for a change in the value of the high-energy cutoff in the mean source
electron spectrum with time. We also show how the construction of the {\it injected} electron
spectrum $F_0(E_0)$ (assuming that Coulomb collisions in a cold target dominate the electron
energetics) is facilitated by the use of higher-order regularization methods.
\end{abstract}
\keywords{Sun: flares, Sun: X-rays, Sun: electron spectrum}
\end{opening}

\section{Introduction}

The hard X-ray spectrum $I(\epsilon)$ (photons~cm$^{-2}$~s$^{-1}$~keV$^{-1}$) in solar flares is
related to the mean electron flux spectrum ${\overline F}(E)$
(electrons~cm$^{-2}$~s$^{-1}$~keV$^{-1}$) through the bremsstrahlung cross-section $Q(\epsilon,E)$
in the defining equation (Brown, Emslie, \& Kontar 2003) for ${\overline F}(E)$, viz.

\begin{equation}
I(\epsilon) = {1 \over 4\pi R^2} \, \, {\overline n} V \int_\epsilon^\infty  \overline F(E) \,
Q(\epsilon,E) \, dE, \label{def}
\end{equation}
where $Q(\epsilon,E)$ is the isotropic bremsstrahlung
cross-section differential in photon energy (cm$^2$~keV$^{-1}$),
$R$ (cm) is the distance to the observer and the mean target
density $\overline n = V^{-1} \int n({\bf r}) \, dV$ (cm$^{-3}$).
Piana et al. (2003) have shown how zero order regularized solution
of Equation~(\ref{def}) for ${\overline F}(E)$ may be obtained
through discretization of the equation and addition of a penalty
term for recovered solutions that exhibit excessive noise in the
solution (cf. Johns \& Lin (1992), who address this issue through
flexible binning of the photon data). Piana et al. (2003) applied
this algorithm to the intense hard X-ray flare of July 23, 2002
(Lin et al., 2003) and derived interesting features in the
recovered ${\overline F}(E)$ spectrum (such as a spectrum
significantly above the forward-fit [Holman et al. 2003],
Maxwellian in the range 20-40~keV, and an apparent dip in the
electron spectrum around 55~keV). This analysis, however, used a
``square'' algorithm, utilizing photon data over the range $10 <
\epsilon < 160$~keV to derive the electron spectrum in the same
energy range; because of the flatness of the spectrum in this
event, they were forced to utilize an extrapolation of the
electron spectrum above 160~keV by a power-law tail to adequately
account for the (significant) contribution of electrons at
energies $E > 160$~keV to the photon emission in the
$[10,160]$~keV range.

As pointed out in the companion paper (Kontar et al. 2004;
hereafter Paper~I), the hard X-ray spectrum over a finite range
$[\epsilon_{\rm min},\epsilon_{\rm max}]$ of photon energies
nevertheless contains considerable information on the {\it
electron} spectrum over a much wider range through the
relation~(\ref{def}). For example, if ${\overline F}(E)$ has an
upper energy cutoff at $E=E_{\rm max} > \epsilon_{\rm max}$, then
one should still find evidence of this upper energy cutoff in the
observed photon spectrum below $\epsilon_{\rm max}$, because of
the requirement that the spectrum tends to zero at $\epsilon =
E_{\rm max}$.  As shown in Paper~I, use of a generalized
(rectangular) regularization method permits a {\it quantitative}
analysis of this high-energy part of the electron spectrum. In
this paper, we therefore apply the analysis technique of Paper~I
to observations of high-resolution hard X-ray spectra obtained
with the {\it Ramaty High Energy Solar Spectroscopic Imager}
(RHESSI) (Lin et al., 2002). We show (\S2) evidence for just such
upper energy cutoffs in the electron spectra in the 26 February,
2002 (about 10:26~UT) solar flare and we discuss the variation of
this cutoff energy with time in \S3. In \S4 we continue the
analysis by presenting the {\it injected} electron flux spectrum
$F_0(E_0)$, under the assumption that the mean source electron
spectrum ${\overline F}(E)$ results from the modification of this
spectrum in a cold collisional target, and we emphasize the
usefulness of higher-order regularization techniques in
constructing this injected spectrum.  In \S5 we present our
conclusions.

\section{Application to RHESSI Data}

\begin{figure}
\begin{center}
\includegraphics[width=89mm]{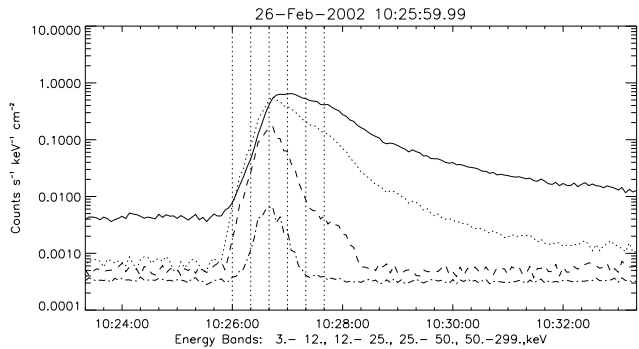}
\end{center}
\caption{Light curves for the 26 February, 2002 event.  The five observational time intervals used
in the analysis are shown by vertical dashed lines.} \label{event}
\end{figure}

\begin{figure}
\begin{center}
\includegraphics[width=54mm]{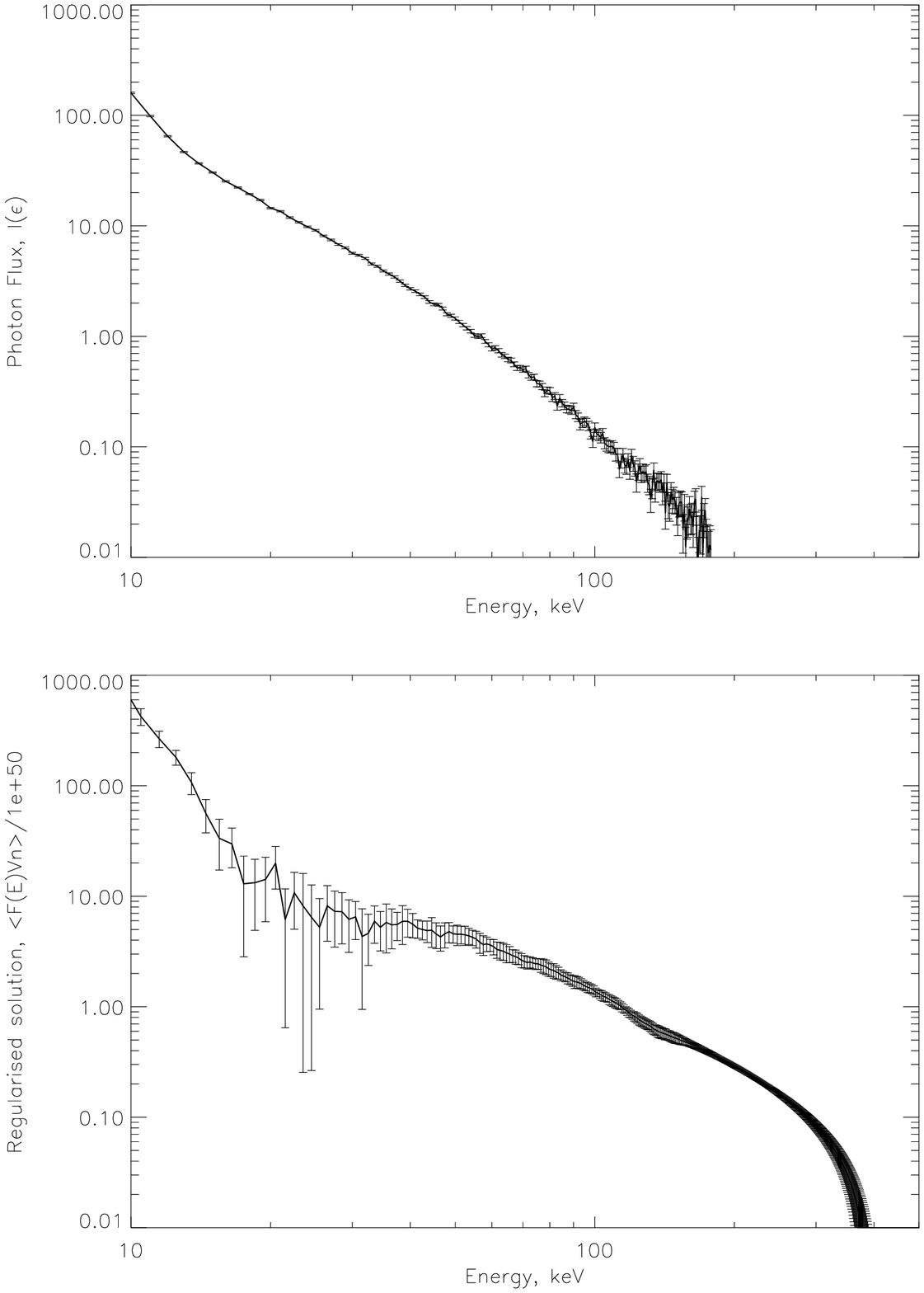}
\includegraphics[width=54mm]{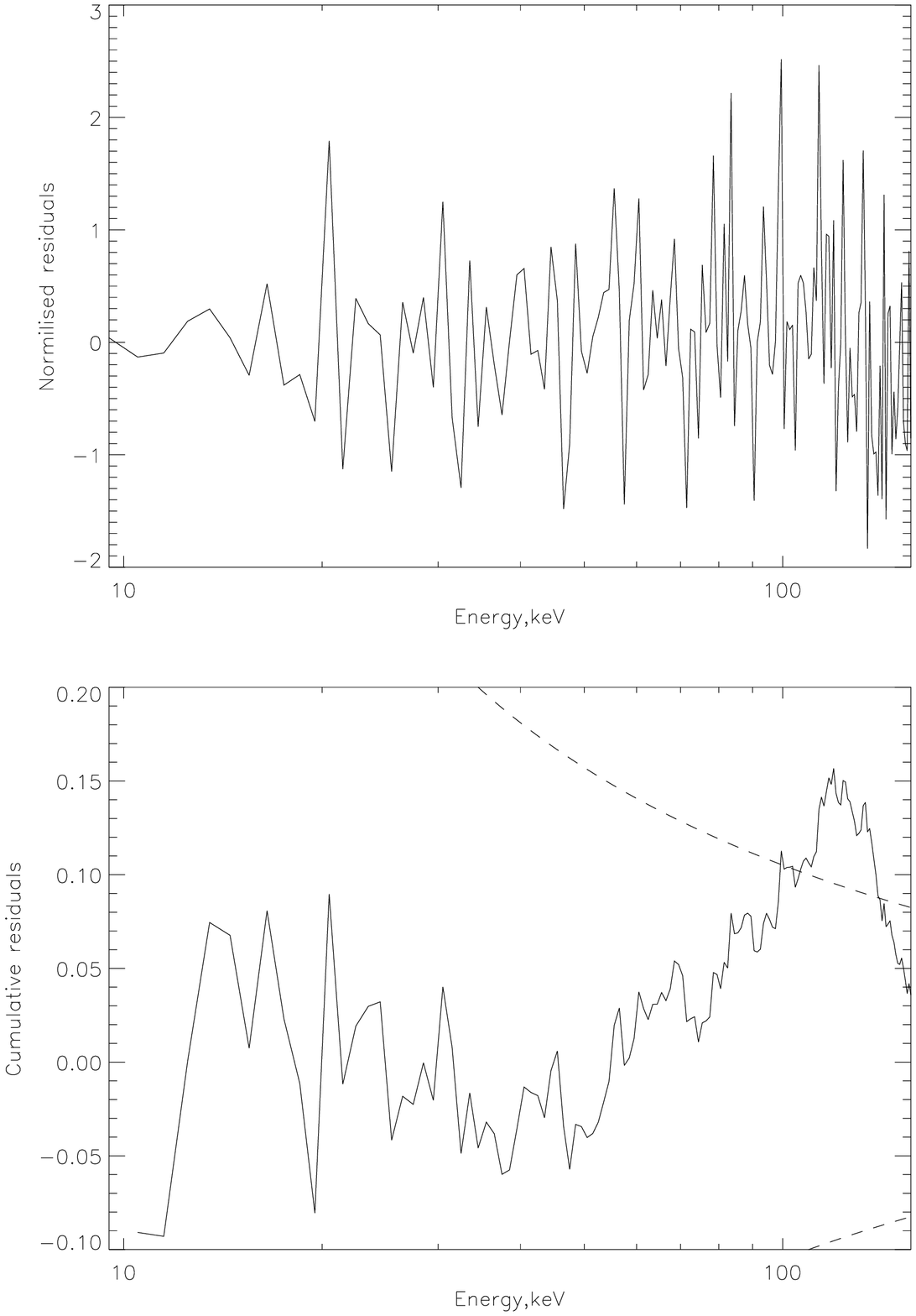}
\end{center}
\caption{{\it Upper left panel:} Photon spectrum for the time
interval from 10:26:20-10:26:40~UT, together with spectrum
reconstructed from the zero order regularized means source
electron spectrum ${\overline F}^{(0)}(E)$, the confidence strip
for which is shown in the {\it lower left panel}. {\it Upper right
panel:} Normalized residuals between the forward-fit photon
spectrum. {\it Lower right panel:} Corresponding cumulative
residuals, compared to the 1$\sigma$ expectation values for a
$\chi^2$ less than 1.} \label{results_zero}
\end{figure}

\begin{figure}
\begin{center}
\includegraphics[width=54mm]{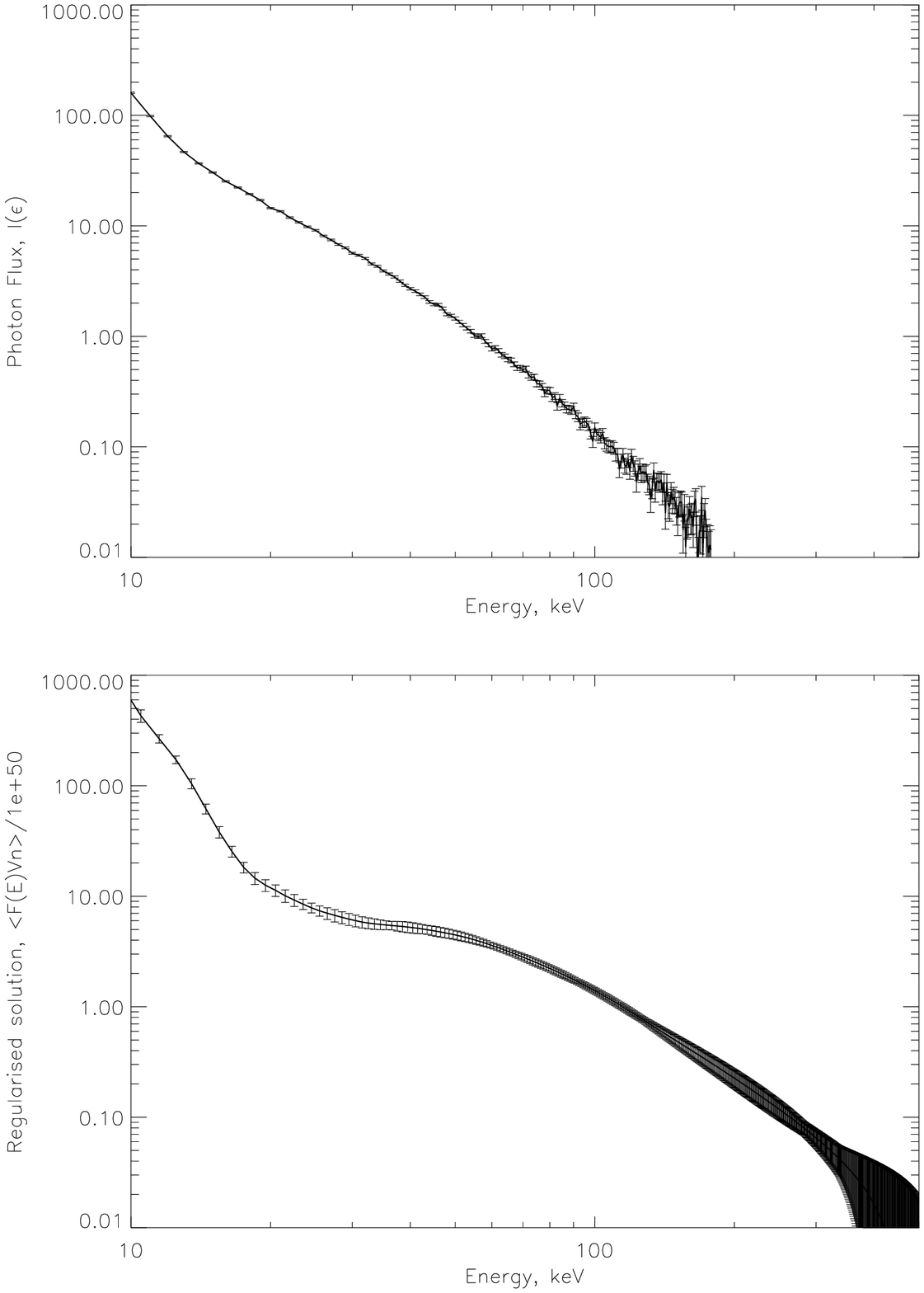}
\includegraphics[width=54mm]{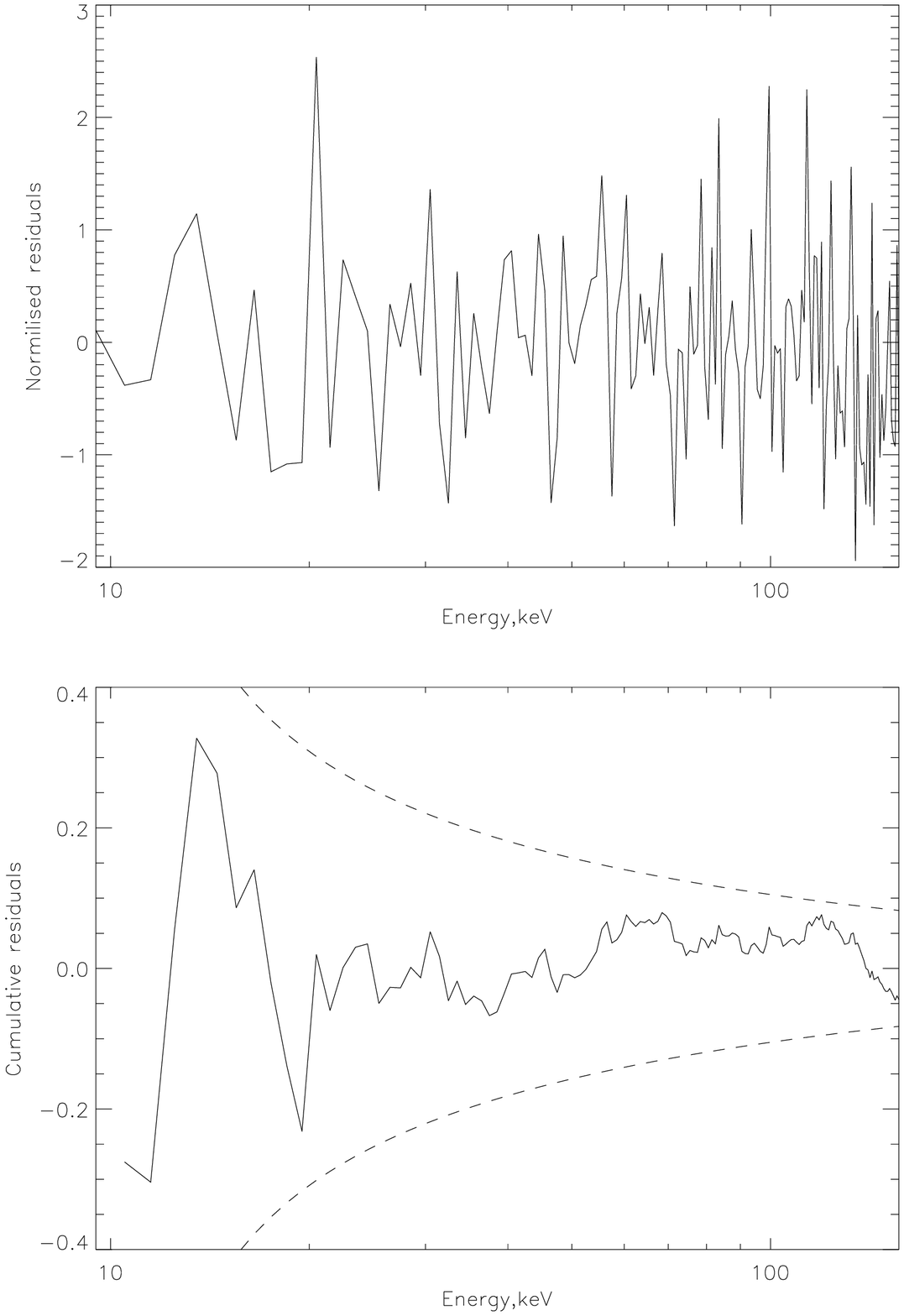}
\end{center}
\caption{As for Figure~\ref{results_zero}, for first order regularization.} \label{results_one}
\end{figure}

\begin{figure}
\begin{center}
\includegraphics[width=54mm]{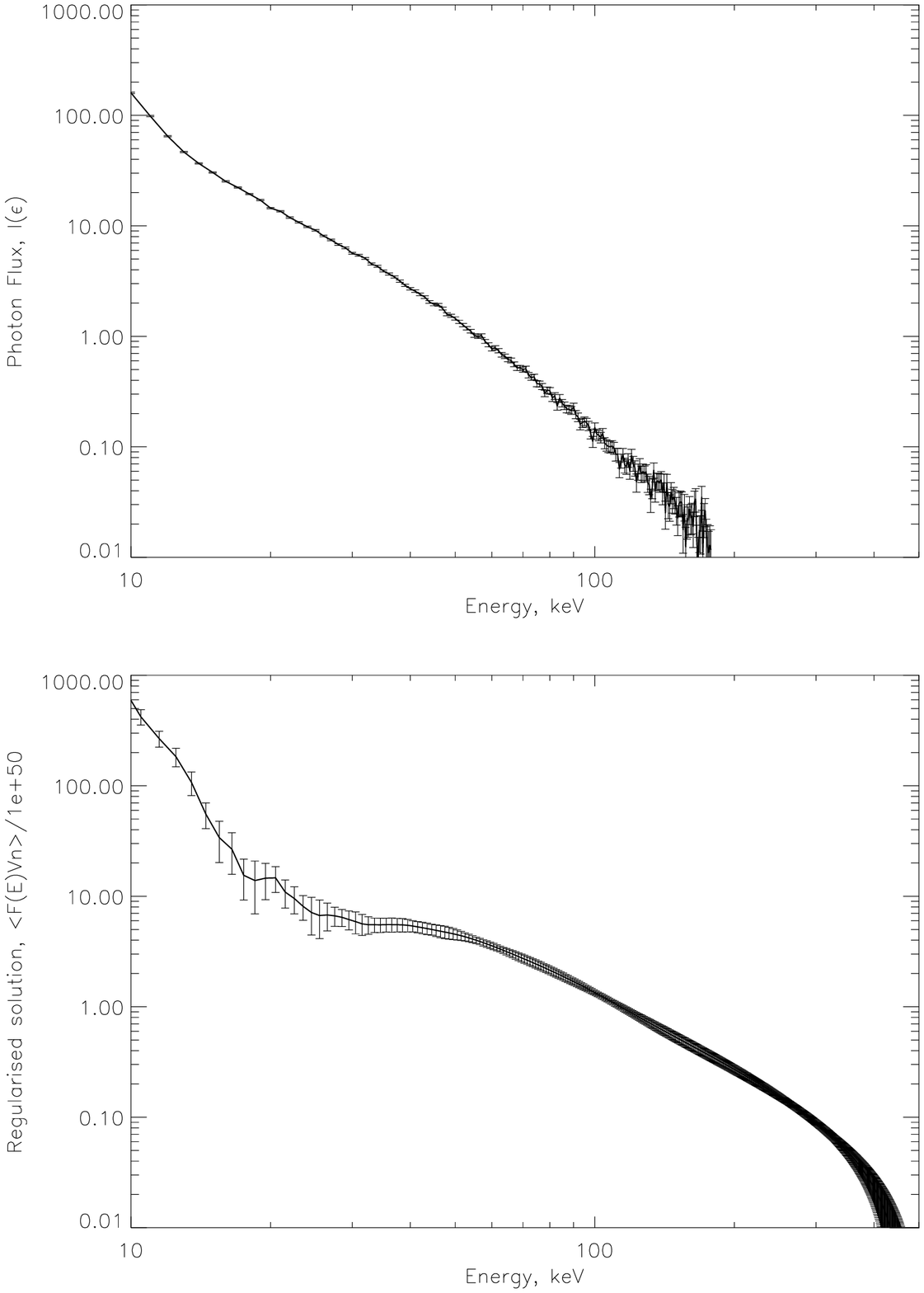}
\includegraphics[width=54mm]{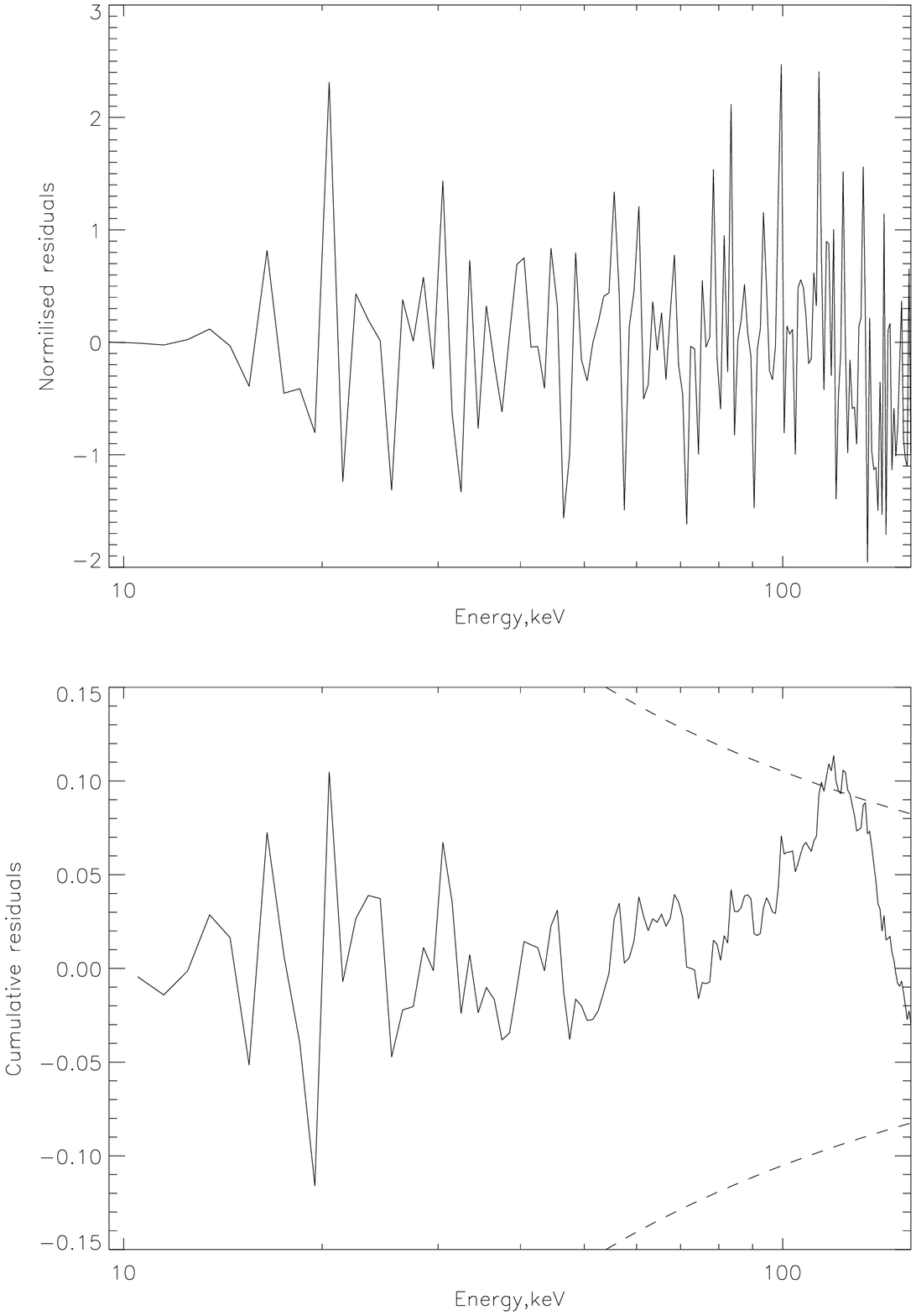}
\end{center}
\caption{As for Figure~\ref{results_zero}, for second order regularization.} \label{results_two}
\end{figure}

\begin{figure}
\begin{center}
\includegraphics[width=89mm]{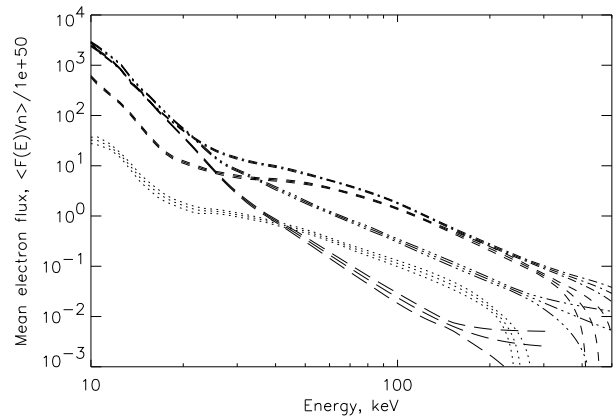}
\end{center}
\caption{Mean source electron spectra for each time interval of
Figure~\ref{event}; Interval 1-dot line, 2 - dash, 3 - dash dot, 4
- dash three dots, 5 - long dash. Thin lines show $1\sigma$
confidence intervals.} \label{dyn_mean}
\end{figure}

\begin{figure}
\begin{center}
\includegraphics[width=89mm]{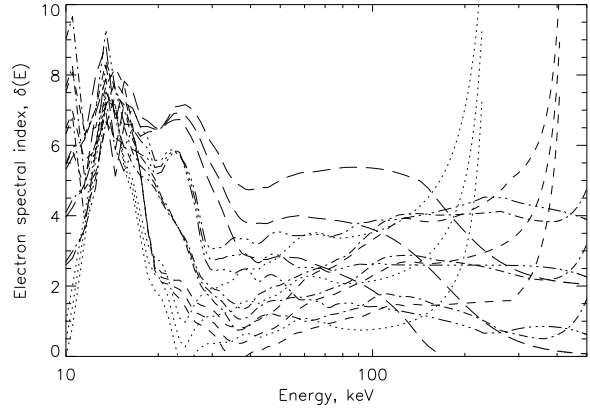}
\end{center}
\caption{Energy variation of the local spectral index $\delta_E =
- d \ln {\overline F}/d \ln E$, obtained using first order
regularization techniques for five time intervals (see Figure 5).
Thin lines show $1\sigma$ error bars.} \label{delta}
\end{figure}

\begin{figure}
\begin{center}
\includegraphics[width=89mm]{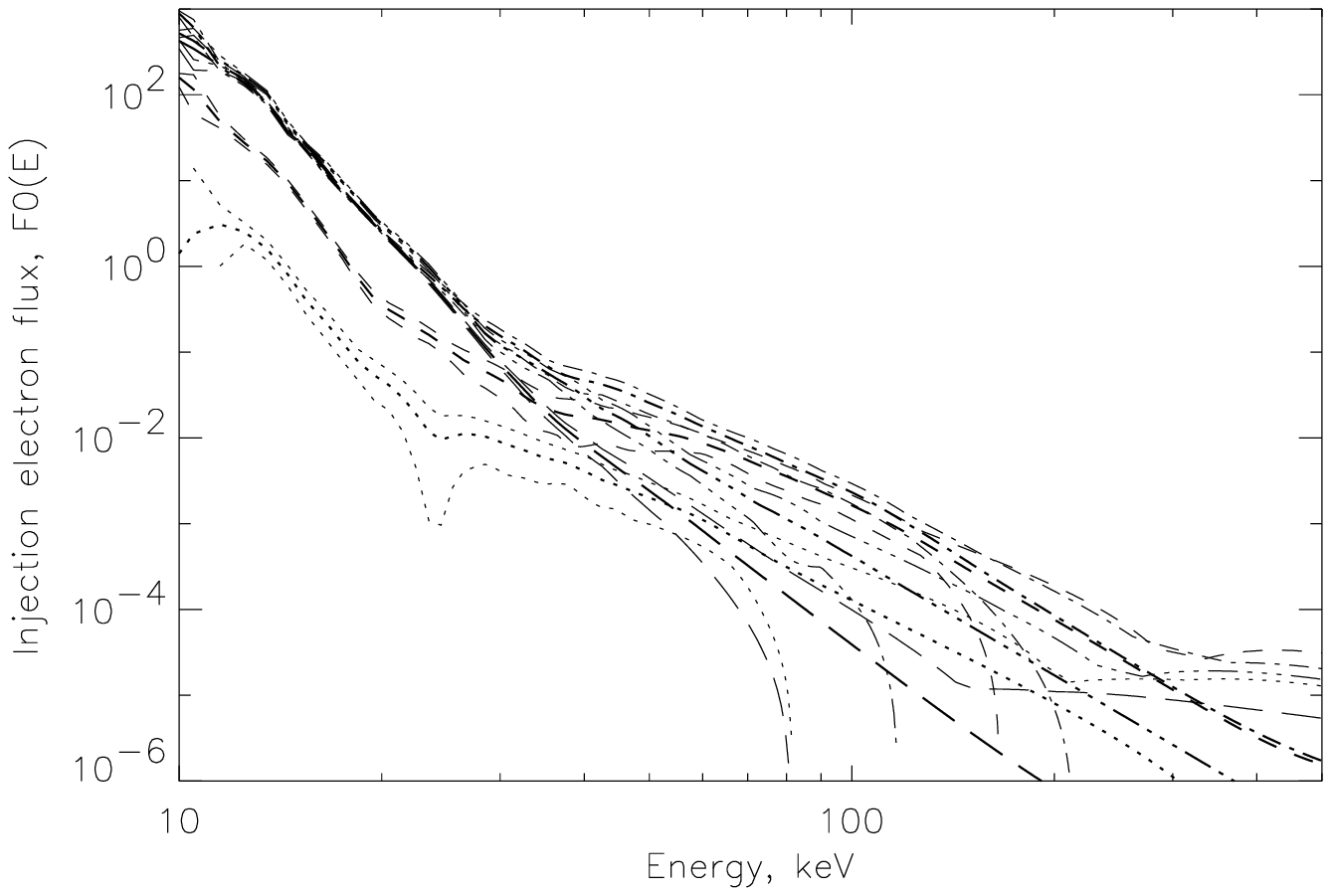}
\end{center}
\caption{Injected electron spectra $F_0(E_0)$ (arbitrary units)
for all 5 time intervals, deduced from the first-order and
equation~(\ref{inj}). Thin lines show $1\sigma$ error bars.}
\label{dyn_inj}
\end{figure}

We selected the GOES M-class hard X-ray event of 26 February, 2002
(start time $\sim$~10:26~UT) for our analysis. This event provides
sufficiently high count rate for a detailed analysis over wide
energy range but does not introduce pulse pile-up issues (Smith et
al, 2002). Figure~\ref{event} shows the light curves for the
photon energy bands $([3-12],[12-25],[25-50],[50-300])$~keV; five
time intervals covering the entire impulsive phase of the flare
are indicated.

RHESSI (Lin et al., 2002) has nine detectors, each designed to have a 1keV resolution in photon
energy. In practice however, detectors 2 and 7 have a poorer resolution than intended (Smith et
al., 2002) and so all results presented here are based only on the seven fully functional front
segments. We used 1 keV energy bins and time bins equal to RHESSI's rotation period (as given for
the time of the flare). This ensured that background and analysis intervals were multiples of the
rotation period eliminating any modulation from the imaging grids.

The top left panel of Figure~\ref{results_zero} shows the photon spectrum for the second time
interval (10:26:20 - 10:26:40~UT), up to a maximum photon energy $\sim 160$~keV for this time
intervals. In the spectrum at energies higher than 160~keV, the uncertainties are comparable to
the signal itself and thus has been ignored (maximum energy in photon spectrum changes from
spectrum to spectrum). The photon data used extended over the range $10 \leq \epsilon \leq
160$~keV, while the electron energy band used in the non-square inversion algorithm extended from
$10 < E < 480$~keV (above 480~keV the solution becomes uncertain). The results are shown as a
confidence interval, i.e. a series of realizations of ${\overline F}(E)$, each corresponding to a
different noise realization of the photon spectrum, with the noise level determined by the
uncertainty in each photon energy bin.

The photon spectrum reconstructed from the mean of this confidence
strip is shown overlaid on the data in the top left panel of
Figure~\ref{results_zero}, while the top right panel of that
Figure shows the residuals (data -- reconstructed spectrum),
normalized to the standard deviation at each energy. The lower
right panel of Figure~\ref{results_zero} shows that the cumulative
residuals (Paper I) mostly fall within the bound, indicating that
the recovered mean source electron spectrum ${\overline F}(E)$
reproduces the observed photon spectrum well, with no systematic
differences between the forward-fitted spectrum and the original
data.

Figure~\ref{results_one} shows the same quantities as
Figure~\ref{results_zero}, for the same photon spectrum, but now
using a {\it first}-order regularization algorithm (see Appendix~A
of Paper~I). Since first-order regularization techniques seek to
minimize the norm of the {\it derivatives} (rather than the
amplitude) of ${\overline F}(E)$, the form of ${\overline F}(E)$
is ``smoother'' than that of ${\overline F}(E)$, particularly
where the slope of ${\overline F}(E)$ changes somewhat abruptly
(e.g., at the transition from ``thermal'' to ``non-thermal'' form
at around 30~keV).  The residuals are similarly well-behaved to
those for the zero order regularization
(Figure~\ref{results_zero}).

Figure~\ref{results_two} shows the same quantities, but the spectrum is smoother in some ranges
and is not for other regions. The lack of smoothness in the solution indicates difficulties to
approximate an electron spectrum with differentiable functions in that region of energies.

Note, that the error bars (Paper I) in all the figures are at
$1\sigma$ level (65\% confidence). Therefore, the errors do not
allow us conclude that the feature near 20 keV seen in
Figure~\ref{results_zero} and Figure~\ref{results_two} is real.

\section{Mean Electron Spectrum}

The mean electron spectrum using zero order regularization ${\overline F}(E)$ for the time
interval (10:26:20-10:26:40~UT) is shown in the Figure (\ref{results_zero}).

It should be noted that the high-energy cutoffs revealed by this
analysis are all at an energy well above the maximum photon energy
sampled. Inversions of the same data set using a binned matrix
method (Johns \& Lin 1992) yield similar results (with much
coarser energy resolution) for the mean source electron spectrum
below 200~keV (Johns-Krull, personal communication) but, by the
intrinsic nature of the matrix inversion method, it cannot provide
unbiased information on the electron spectrum above $E =
\epsilon_{\rm max}$.

We note that both zero order and first order ${\overline F}(E)$
curves become much smoother above $\epsilon_{\rm max}$, because of
the lack of (noisy) data at such high energies. This result
illustrates convincingly the power of the regularization technique
to use subtle features in the photon spectrum to provide valuable
information on the source electron spectrum over a wide range of
energies (Paper I).

\section{Injected Electron Spectra}

The spectra ${\overline F}(E)$ defined by Equation~(\ref{def}) are {\it mean source electron
spectra}, and so represent (density-weighted) averages of the electron flux spectra over the
entire source (Brown, Emslie, \& Kontar 2003). Emslie (2003) has provided a formula (his Equation
[5]) linking the {\it injected} electron spectrum $F_0(E_0)$ to ${\overline F}(E)$, for a given
energy loss model for the electrons.  If we approximate the target by a fully ionized ``cold''
target, then we recover Emslie's Equation (6), viz. (see Brown \& Emslie 1988)

\begin{equation}
F_0(E_0) = {{\overline n} V \over A} \, K \, {\overline F(E_0) \over E_0^2} \left ( 1 - {d \ln
{\overline F} \over d \ln E} \right )_{E=E_0}. \label{inj}
\end{equation}
The appearance of the derivative $d \ln {\overline F}/d \ln E$ in
this expression means that in calculating injected electron
spectra $F_0(E_0)$, it is important to accurately estimate not
only ${\overline F}(E)$, but also its logarithmic slope. Hence
regularized solutions that seek to minimize variations in this
slope from point to point will produce smoother forms of
$F_0(E_0)$.

Figure~\ref{delta} shows the variation of the local spectral index
$\delta_E = - d \ln {\overline F}/d \ln E$ with $E$. Because the
second-order regularization method is specifically designed to
minimize fluctuations in the second derivative, the variation of
the first derivative (and so $\delta_E$) is much smoother than for
the other two orders of regularization. We therefore expect that a
second-order regularization method will produce the smoothest
forms of the injected electron spectrum $F_0(E_0)$.

Figure~\ref{dyn_inj} shows the forms of the injected spectrum
$F_0(E_0)$ in arbitrary units for all five time interval using
first order regularization methods.  As expected, the second order
solution $F_0(E_0)$ shows the most ``reasonable'' behavior. Not
surprisingly, the injected electron spectra show high-energy
cutoffs at values similar to that of the corresponding mean source
electron spectra.

\section{Summary and Discussion}

The most important result of this study is that hard X-ray spectra
over a finite range of photon energies still carry vital
information on the responsible electron spectrum at electron
energies substantially higher than the maximum photon energy
observed, and that this information can be extracted
quantitatively using a non-square Tikhonov regularization
technique (Paper I). Further, use of higher-order regularization
techniques can produce mean source electron spectra with forms
suitable for further analysis, such as the calculation of the
electron spectrum, injected into a collisional thick target, that
produces the inferred mean source electron spectrum.

Application of this technique to a flare on February 26, 2002 has
shown that the maximum accelerated electron energy rises and falls
with time after the peak of the event, concurrent with a growing
low-energy thermal component of the hard X-ray emission.

This reduction in the maximum accelerated electron energy may be associated with the changing
atmospheric conditions evidenced by the enhanced thermal emission at these later times; the
increase in soft X-ray emission may be connected with a larger coronal density which may act to
suppress electron acceleration to high energies. Electrons subjected to an applied electric field
${\cal E}$ obey the equation of motion

\begin{equation}
{dE \over ds} = e \, {\cal E} \left \{ 1 - \left ( {v_t \over v} \right )^2 { {\cal E}_D \over
{\cal E} } \right \}, \label{motion}
\end{equation}
where $e$ is the magnitude of the electronic charge, $v_t$ is the
electron thermal velocity and ${\cal E}_D \simeq 10^{-7} n/T$
(V~cm$^{-1}$) is the Dreicer field (Dreicer, 1959).  They
therefore suffer runaway accleration above a critical velocity

\begin{equation}
v_{\rm crit} = v_t \sqrt{{\cal E}_D/{\cal E}}. \label{dreic}
\end{equation}
The electrons that suffer runaway acceleration emerge from an
acceleration region of length $L$ with an energy given by the
solution of Equation~(\ref{motion}).  As the density $n$
increases, commensurate with the increased emission measure
evidenced by the enhanced thermal emission at later time intervals
(Figure~\ref{dyn_mean}), the value of the Dreicer field is
correspondingly increased, the ratio ${\cal E}_D/{\cal E}$
increases and, by Equation~(\ref{motion}), the net force on each
electron, and hence its emergent energy (``injection energy'') is
decreased.

\subsection*{Acknowledgment} This work was supported by NASA's Office of Space Science through Grants NAG5-207745,
by a PPARC rolling grant, and by a collaboration grant from the Royal Society.


\end{article}
\end{document}